\documentclass[11pt]{article}
\usepackage{amsmath}
\usepackage{dsfont}
\usepackage{tikz}
\usepackage{tensor}
\usepackage{simplewick}
\usepackage{amssymb}
\usepackage{tensor}
\usepackage{titling}
\usepackage{lipsum}
\usepackage{physics}
\usepackage{color}   
\usepackage{hyperref}
\hypersetup{
    colorlinks=true,
    linktoc=all,     
    linkcolor=black, 
}
\usepackage{soul}
\usepackage{hyperref}
\usepackage{authblk}
\usepackage{sectsty}
\usepackage[utf8]{inputenc}
\usepackage{geometry}
\sectionfont{\fontsize{11}{15}}
\numberwithin{equation}{section}
\usepackage{fancyhdr}

\title{Double Field Theory for Generalized $\lambda-$deformation}
\author{Parita Shah\footnote{pshah3@albany.edu}}
\affil{Department of Physics\\University at Albany (SUNY)\\
Albany, NY 12222, USA }
\date{}
\begin{document}
\maketitle
\begin{abstract}
We embed the geometries of the generalized $\lambda-$deformations into the framework of the Double Field Theory.
\end{abstract}
\newpage
\tableofcontents
\newpage
\section{Introduction}
Generalized dualities such as abelian, non-abelian T-duality and Poisson Lie (PL) T-dualities have been the objects of interests in the area of holography as they can be used as solution generating techniques within supergravity. A doubled worldsheet formalism was introduced earlier to make these abelian T-dualities manifest in the construction \cite{Tseytlin1990DualitySF}. This formalism constructed a $\sigma-$model on a double dimensional target space such that it reduced to the known standard $\sigma-$model when restricted to half of the coordinates. Similarly, a doubled target space formulation called the Double Field Theory (DFT) was introduced in \cite{2009, hull1003} for the case of abelian T-dualities. This work was further extended for other generalized dualities as well \cite{1995,1996,hull2009non, demulder2019doubled, 17new}. To collect all these generalized cases, a neat doubled worldsheet formalism was introduced known as the $\mathcal{E}-$models. These $\mathcal{E}$-models have included not only PL-dualisable $\sigma-$models but also $\eta-$ and $\lambda-$deformed models \cite{klimvcik2015eta}. Having an $\mathcal{E}-$model can correlate to defining a DFT on a 2D-dimensional $\mathbb{D}$ with invariance under 2D-diffeomorphisms.
\par Complicated calculations in integrable deformations can be simplified by the use of PL-symmetry, making the study of $\mathcal{E}-$models and DFTs useful \cite{demulder2019invitation,thompson2019introduction}. DFT enables us to include the integrable models under a single umbrella, as was shown in \cite{borsato2021supergravity, 21new}. In terms of DFT, various integrable models can be regarded as mere examples or special cases of these theories. The DFT formulation manifests abelian T-duality at the level of low energy effective action by using a doubled coordinate using the coordinates that span the usual D-dimensional target space as well as the coordinates of the T-dual. In the working of DFT, T-duality maps, diffeomorphisms and B-field gauge transformations are all viewed as $O(D,D)$ transformations, where the DFT action remains invariant under the group $O(D,D)$. Imposing the strong constraint on the DFT fields, i.e. demanding that they depend only on the D-dimensional target space coordinates, the low-energy effective action for strings is recovered. The $O(D,D)$ scalar in DFT acts similar to the dilaton $\phi$ and the generalized $O(D,D)$ invariant metric $\mathcal{H}_{MN}$ includes the information about the metric and the B-field. By the virtue of DFT, T-dualities can also be visualized in non-isometric directions \cite{Hull_2005,hull2009non,Andriot_2012,20192}.
\par  By viewing the $\lambda-$model in the framework of $\mathcal{E}-$models, \cite{klimvcik2015eta} shows how we are able to see beyond the possible values of $\lambda$ which were set to the range $0\leq \lambda\leq 1$. This indicates how working in the doubled framework of $\mathcal{E}-$models/DFT more underlying subtleties can be brought to light. \cite{demulder2019doubled} gives a target space description to the worldsheet $\mathcal{E}-$models within the DFT for the usual $\lambda-$deformed model on $\mathbb{D}/\tilde{H}$. \cite{borsato2021supergravity} follows the work of \cite{demulder2019doubled} and uses the generalized Scherk-Schwarz ansatz to find the generalized frame field and extends the work to show how the same algebra can also cover the asymmetrical $\lambda-$deformation model. \cite{borsato2021supergravity} also discusses solution-generating techniques which are used to construct ways to find a set of metric, Kalb-Ramond field, and, dilaton (Neveu-Schwarz Neveu-Schwarz (NSNS) fields) which would also form a supergravity solution considering a known set of NSNS fields which are themselves a supergravity solution. This is performed by going to the doubled fields of DFT.
\par It was an interesting direction to find ways to deform an integrable non-linear $\sigma-$model such that the integrability would still hold. $\lambda-$deformed $\sigma-$model, Yang-Baxter (YB) $\sigma-$model \cite{20091} are some of the well-studied examples of such integrable deformed models. Few years later after $\lambda-$model was introduced \cite{2014}, Sfetsos et. al. \cite{sfetsos2015generalised} extended the study to include non-trivial $\lambda$ deformations, in the form of matrices rather than a number or parameter and formulated the generalized $\lambda-$deformation model\footnote{Another extension of the $\lambda-$deformation was proposed recently in \cite{20212}.}. These generalized models elegantly not only included the usual previously known $\lambda-$model but also included the YB deformed models as a special case of the same generalized $\lambda-$model.
\par In this paper,we establish a doubled formalism in terms of the worldsheet description given by an $\mathcal{E}-$model as well as transform this to give a target space description in terms of a DFT for the generalized $\lambda-$deformation model. In section \ref{2}, we recollect the known background and set up the general conventions and that are to be followed in the rest of the paper. In section \ref{3}, we use the aforementioned procedure for the case of a generalized $\lambda-$deformation to get the corresponding $\mathcal{E}-$model and DFT description. In section \ref{4}, we follow the work of \cite{borsato2021supergravity} to embed the asymmetrical generalized $\lambda-$deformation model into DFT. We conclude in section \ref{5} by briefly summarizing the work and discussing some interesting directions. Some technical details are presented in appendices. Appendix \ref{A} introduces the conventions used in this paper in a detailed manner. Appendix \ref{app eta} reviews the relationship between the bi-Yang-Baxter model and the generalized $\lambda$-deformation which makes the DFT embedding of section \ref{3} applicable to the generalized $\eta$-deformation as well. Appendix \ref{B} describes the basics of an $\mathcal{E}-$model, and Appendix \ref{C} assists the work of section \ref{3} by giving the details on the derivation of the $\mathcal{E}-$map, projection operator $\mathcal{P}_m(\mathcal{E})$ as well as the generalized metric $\mathcal{H}_{AB}$.
\section{Review}\label{2}
We begin by gathering the results and conventions known and used in some of the previous works done in \cite{klimvcik2015eta,demulder2019doubled}. We recall the construction of a double field theory as well as its corresponding worldsheet description given by $\mathcal{E}-$model. We further revisit the generalized $\lambda-$model which will be used in the next section for contructing a doubled framework of the deformed model.
\subsection{Double Field Theory construction}
The doubled theory considers a real Lie algebra $\mathcal{D}$ of group $\mathbb{D}$ with dimension $2D$ equipped with an invariant, non-degenerate, symmetric inner product $(.,.)_{\mathcal{D}}$. The generators for the algebra are given by the basis $\mathbb{T}_A$ such that,
\begin{equation}\label{commutation relation}
\left[\mathbb{T}_{A}, \mathbb{T}_{B}\right]=\mathbb{F}_{A B}{ }^{C} \mathbb{T}_{C}, \quad\left(\mathbb{T}_{A}, \mathbb{T}_{B}\right)_{\mathcal{D}}=\eta_{A B}
\end{equation}    
The $\eta_{AB}$ and the inverse $\eta^{AB}$ are used to lower and raise the indices respectively. A Double Field Theory (DFT) is an $O(D,D)$ invariant theory defined on this group $\mathbb{D}$. DFT has been used to embed non-linear $\sigma$-models and understand non-trivial T-dualities as transformations which combine a theory on coordinates $x^i$ and its dual theory defined on $\tilde{x}_{\tilde{i}}$ where $i,\tilde{i}=1,\ldots,D$. Therefore, the group $\mathbb{D}$ is spanned by coordinates $X^I=(\tilde{x}_{\tilde{i}},x^i)$ where $I=1,\ldots,2D$. The DFT is identified by a generalized metric $\mathcal{H}_{AB}$ (may depend on $X^I$), generalized dilaton and other corresponding generalized Ramond-Ramond (RR) fields. The DFT action constructed from these fields is invariant under conventional 2D diffeomorphisms and generalized diffeomorphisms. Here, consistency is ensured by implementing section condition that demands the fields to depend only on the physical $x^i$ coordinates. In order to incorporate the embedding of a non-linear $\sigma-$model and the section condition, a maximally isotropic subalgebra with respect to $(.,.)_{\mathcal{D}}$ is considered $\tilde{\mathcal{H}}\subset\mathcal{D}$ corresponding to the subgroup $\tilde{H}\subset\mathbb{D}$. It has been known\cite{hull2009non} that such a set up implies that there exists a non-linear $\sigma$-model on the target space $\mathbb{D}/\Tilde{H}$. The basis is now written as $\mathbb{T}_A=(\Tilde{T}^a,T_a)$, where $\Tilde{T}^a$ are the generators of $\Tilde{{H}}$ and $T_a$ are the generators for $\mathbb{D}/\Tilde{H}$. Therefore, in this basis, 
\begin{equation}\label{eta}
\eta_{AB}=\begin{pmatrix}
0&\delta^a{}_b\\
\delta_a{}^b&0
\end{pmatrix}
\end{equation}
It is not necessary for $\mathbb{D}/\Tilde{H}$ to be either a group manifold or a symmetric space, however, in order to relate this formalism with the deformed models that will be the case. The DFT is now identified by $x^i$ dependent generalized metric $\widehat{\mathcal{H}}_{\hat{I}\hat{J}}(x^i)$, generalized dilaton $\hat{d}$ and other modified fields. The information of metric $G$ and $B-$field of the physical theory embedded in the doubled framework of DFT can be seen in this generalized metric,
\begin{equation}
\widehat{\mathcal{H}}_{\hat{I} \hat{J}}(x)=\left(\begin{array}{cc}
G^{-1} & -G^{-1} B \\
B G^{-1}& G-B G^{-1} B
\end{array}\right)_{\hat{I}\hat{J}}
\end{equation}
The hatted indices indicate that the objects depend only on the coordinates $x^i$ and not on $\tilde{x}_{\tilde{i}}$. The two generalized metrics $\mathcal{H}_{AB}$ and $\widehat{\mathcal{H}}_{\hat{I}\hat{J}}(x)$ are related by the coordinate dependent generalized frame fields $\tensor{{\widehat{E}}}{^A_{\hat{I}}}$ whose existence is ensured by the way the target space $\mathbb{D}/\tilde{H}$ is constructed \cite{demulder2019doubled},
\begin{equation}
\widehat{\mathcal{H}}_{\hat{I} \hat{J}}(x)=\widehat{E}^{A}{}_{\hat{I}}(x) \mathcal{H}_{A B} \widehat{E}^{B}{ }_{\hat{J}}(x)
\end{equation}
The generalized metric $\mathcal{H}_{AB}$ is used to describe the worldsheet description corresponding to DFT known as the $\mathcal{E}-$model. The $\mathcal{E}-$model is a worldsheet theory defined by a real linear map $\mathcal{E}:\mathcal{D}\rightarrow\mathcal{D}$ such that it is self-adjoint under the inner product $(.,.)_{\mathcal{D}}$ and idempotent, which implies,
\begin{equation}\label{cond for E map}
(\mathcal{E}x,y)_{\mathcal{D}}=( x,\mathcal{E}y)_{\mathcal{D}} ~~~\forall~x,y\in \mathcal{D}~~~~~~~~~~~~~~~~~~\mathcal{E}^2x=x   ~~~\forall~x\in \mathcal{D}
\end{equation}
Note that, in terms of the basis, $\mathcal{E}(\mathbb{T}_A)=\mathcal{E}_A{}^B\mathbb{T}_B$. The map $\mathcal{E}$ is parametrized in terms of a generalized metric $\mathcal{H}$,
\begin{equation}\label{cond for E and H}
\mathcal{E}_A{}^B=\mathcal{H}_{AC}\eta^{CB}    
\end{equation}
The generalized metric is symmetric and also obeys the relation,
\begin{equation}
\mathcal{H}_{A B}=\mathcal{H}_{B A}, \quad \mathcal{H}_{A C} \eta^{C D} \mathcal{H}_{D B}=\eta_{A B}
\end{equation}

\par The following action of the $\mathcal{E}-$model describes the corresponding non-linear $\sigma-$model defined by the $\mathcal{E}-$map on the target space $\mathbb{D}/\tilde{H}$,
\begin{equation}\label{action DFT}
\begin{aligned}
S_{\mathbb{D} / \tilde{H}} &=\tilde{k} S_{W Z W}[f]-\frac{\tilde{k}}{\pi} \int d \sigma d \tau\big(\mathcal{P}_f(\mathcal{E})\left(f^{-1} \partial_{+} f\right), f^{-1} \partial_{-} f\big)\\
S_{W Z W}[f] &=\frac{1}{2 \pi} \int d \sigma d \tau\big(f^{-1} \partial_{+} f, f^{-1} \partial_{-} f\big)+\frac{1}{24 \pi} \int_{\text{bulk}}\big( f^{-1} d f,\left[f^{-1} d f, f^{-1} d f\right]\big)
\end{aligned}
\end{equation}
Here, for an element ${g}(X^I)\in\mathbb{D}$, it is decomposed as $g(X^I)=\tilde{h}(\tilde{x}_{\tilde{i}})f(x^i)$ such that $\tilde{h}(\tilde{x}_{\tilde{i}})\in \tilde{H}$ and $f(x^i)\in \mathbb{D}/\tilde{H}$. The $\mathcal{P}_f(\mathcal{E})$ is a projection operator constructed as follows,
\begin{equation}\label{cond for projector}
\mathcal{P}_f(\mathcal{E}):\mathcal{D}\rightarrow\mathcal{D}~~~~~~~~~~~~
\operatorname{Im}\mathcal{P}_f(\mathcal{E})=\tilde{\mathcal{H}}~~~~~~~~~ \operatorname{Ker}\mathcal{P}_f(\mathcal{E})=(\mathbb{I}+\operatorname{Ad}_{f^{-1}}\mathcal{E}\operatorname{Ad}_f)\tilde{\mathcal{H}}
\end{equation}
The $\mathcal{E}-$model is constructed to obtain the generalized metric $\mathcal{H}_{AB}$ for the corresponding embedded model. Non-linear $\sigma-$models for $\eta-$ and $\lambda-$deformations have been established as an $\mathcal{E}-$model and hence as a DFT earlier in \cite{demulder2019doubled, klimvcik2015eta}. An important step in writing a $\sigma-$model in the doubled framework is to construct an appropriate $\mathcal{E}-$map, as per the given conditions, which would correctly reproduce the action of the desired model from the $\mathcal{E}-$model action (\ref{action DFT}). In the next section, this background setup will be used to build a similar structure of the $\mathcal{E}-$model and DFT for the specific case of the generalized $\lambda-$model.
\subsection{Generalized $\lambda-$model}
Here we briefly review the generalized $\lambda-$deformed model  which will be embedded in a doubled formalism in the next section. The integrable generalized $\lambda-$deformed model was constructed \cite{sfetsos2015generalised} to treat non-trivial matrix deformations $\lambda$ as compared to the model where $\lambda$ was considered to be a number \cite{2014}. The known integrable principal chiral model (PCM) and Wess-Zumino-Witten (WZW) model are combined to obtain the generalized $\lambda-$model. A PCM model and a WZW model defined on group $G$ are considered as follows,
\begin{equation}\label{PCM}
S_{PCM}(\hat{g})=\frac{1}{2\pi}\int d^2\sigma \hat{F}_{ab}v^a_+(\hat{g})v^b_-(\hat{g}) ~~~~~~~~~~~~~~~~~~~~\hat{g}\in G  
\end{equation}
\begin{equation}\label{WZW}
S_{WZW}(g)=\frac{k}{4\pi}\int_{\sum}d^2\sigma v^a_+v^b_--\frac{k}{24\pi}\int_{\text{bulk}}f_{abc}v^a\wedge v^b\wedge v^c    ~~~~{g}\in G 
\end{equation}
where, the $\hat{F}_{ab}$ is an arbitrary matrix deformation in the PCM and the adjoint action and left-right invariant Maurer-Cartan forms are given below where $t^a$ are the generators and $f_{ab}{}^c$ is the structure constant of algebra $\mathcal{G}$ for corresponding group $G$  
\begin{equation}
D_{a b}=\operatorname{Tr}\left[t^{a} g t^{b} g^{-1}\right] \quad e_{\pm}^{a}=i \operatorname{Tr}\left[t^{a} g^{-1} \partial_{\pm} g\right] \quad v_{\mu}^{a}=D_{a b} e_{\mu}^{b}
\end{equation}
The two actions are added and the following subgroup $H$ of symmetry is gauged by introducing a gauge field $A\rightarrow h^{-1}Ah+h^{-1}dh$,
\begin{equation}
\hat{g}\rightarrow\hat{g}h~~~~~~~~~~~~~g\rightarrow h^{-1}gh~~~~~~~~~h\in G
\end{equation}
The gauge is further fixed by $\hat{g}=\mathbb{I}$ and the gauge fields are integrated out to give the action for generalized $\lambda-$model,
\begin{equation}\label{lambda model}
\begin{aligned}
S_{\lambda}(g)=S_{WZW}(g)&+\frac{k}{2\pi}\int d^2\sigma e^a_+(\lambda^{-1}-D)_{ab}^{-1}v^b_-\\
\text{where,~~~~~~~~~~~}\lambda^{-1}&=\frac{1}{k}(\hat{F}+k\mathbb{I}) 
\end{aligned}
\end{equation}
It should be noted that this model is integrable only for particular choices of the matrix $\hat{F}_{ab}$ and hence for specific $\lambda$ \cite{sfetsos2015generalised}. In particular, in the Appendix \ref{app eta}, we review the incorporation of the integrable bi-Yang-Baxter model into the framework of the generalized $\lambda-$deformation, making the results of this article on embedding of integrable models into DFT formalism applicable to the bi-Yang-Baxter deformation.\footnote{We would like to thank the referee for suggesting this addition.}  In the next section, we develop a DFT which embeds the system (\ref{lambda model}) in a doubled formalism, however, we treat $\lambda$ as an arbitrary matrix in the construction without assuming any specific choice.
\section{Generalized $\lambda$-model in doubled formalism}\label{3}
In this section, we construct the non-linear $\sigma-$model of the generalized $\lambda-$deformation in the doubled framework of $\mathcal{E}-$model and DFT. We follow a procedure similar to \cite{klimvcik2015eta,demulder2019doubled} where $\eta-$ and $\lambda-$deformed models were formulated as $\mathcal{E}-$models. However, we employ this procedure here for a generalized $\lambda-$model where $\lambda$ is considered a matrix. This results in the corresponding $\mathcal{E}-$model that is further used to determine the generalized metric $\widehat{\mathcal{H}}_{\hat{I}\hat{J}}(x)$ which identifies the DFT for generalized $\lambda-$model. Therefore, we first develop the $\mathcal{E}-$model in section \ref{3.1} and obtain the DFT generalized metric and address the embedding of the generalized $\lambda-$model in section \ref{3.2}.
\subsection{$\mathcal{E}-$model for generalized $\lambda$-model}\label{3.1}
To establish the $\mathcal{E}-$model first, we begin by constructing an $\mathcal{E}-$map for the generalized $\lambda-$deformed model. Our main aim in this section is to obtain a projection operator $P_f(\mathcal{E})$ using the $\mathcal{E}-$map, such that upon substitution the $\mathcal{E}-$model action (\ref{action DFT}) gives the known action \cite{2014, sfetsos2015generalised} of the generalized $\lambda-$deformation model. The generalized $\lambda-$model action from (\ref{lambda model}) is rewritten in terms of the group element as follows (refer Appendix \ref{A} for details),
\begin{equation}\label{lambda action}
\begin{aligned}
S_{\lambda}(g)&=S_{\text{WZW}}(g)+ \int d \xi^{+} d \xi^{-}\left(\partial_{+} g g^{-1},({\lambda}^{-1}- \mathrm{Ad}_{g^{-1}}\right)^{-1} g^{-1} \partial_{-} g)\\
S_{\text{WZW}}(g)&=\frac{1}{2} \int d \xi^{+} d \xi^{-}\left(g^{-1} \partial_{+} g, g^{-1} \partial_{-} g\right)+\frac{1}{12} \int d^{-1}\left(d g g^{-1},\left[d g g^{-1}, d g g^{-1}\right]\right) 
\end{aligned}
\end{equation}
\par Following \cite{demulder2019doubled,borsato2021supergravity,klimvcik2015eta}, we consider the Drinfeld double $\mathcal{D}$ and the direct sum $\mathcal{G}\oplus\mathcal{G}$, where $\mathcal{G}$ is the real Lie algebra for the group $G$. The generators for $\mathcal{G}$ are given by $t_a$ such that $\kappa_{ab}=\langle t_a,t_b\rangle$. For the coset $M=\mathbb{D}/\tilde{H}$, the corresponding maximally isotropic subalgebra $\tilde{\mathcal{H}}$ is chosen to be the diagonal subalgebra, $\mathcal{G}_{\text{diag}}$ embedded in $\mathcal{G}\oplus\mathcal{G}$ by the map $X\rightarrow\frac{1}{\sqrt{2}}\{X,X\}$ for $X\in \mathcal{G}$. On the other hand, the anti-diagonal subspace ${G}_{\text{anti-diag}}$ is embedded as $X\rightarrow\frac{1}{\sqrt{2}}\{X,-X\}$. Here, the isotropic subspace $G_{\text{anti-diag}}$, ${M}=\mathbb{D}/\tilde{H}$ does not form a subgroup however is still a symmetric space allowing us to construct a non-linear $\sigma-$model on it.
\par Instead of working on the double $\mathcal{D}$, we work on the direct sum $\mathcal{G}\oplus\mathcal{G}$ \cite{klimvcik2015eta}. A map between the algebras $\mathcal{D}$ and $\mathcal{G}\oplus\mathcal{G}$ is required since the action (\ref{action DFT}) is constructed using the algebra $\mathcal{D}$. It is known that for each $\lambda$ there exists an isomorphism $\Phi_{\lambda}: \mathcal{D}\rightarrow \mathcal{G} \oplus \mathcal{G}$ such that it preserves the inner product on $\mathcal{D}$ and $\mathcal{G}\oplus\mathcal{G}$ up to a constant factor dependent on $\lambda$ where, the inner product for the double $\mathcal{D}$ is given by \cite{klimvcik2015eta},
\begin{equation}\label{D inner product}
\left(x_{1} \dot{+} x_{2}, y_{1} \dot{+} y_{2}\right)_{\mathcal{D}}:=\left(x_{2}, y_{1}\right)+\left(x_{1}, y_{2}\right)
\end{equation}
elements of the type $(x_1\dot{+}x_2)$ belong to $\mathcal{D}$, whereas, the elements of the direct sum $\mathcal{G}\oplus\mathcal{G}$ are written as $\{\alpha,\beta\}$, with $x_i, y_i\in\mathcal{G}$ and the inner product for $\mathcal{G}\oplus\mathcal{G}$ is as follows,
\begin{equation}\label{GG inner product}
\left.\left(\left\{\alpha_{1}, \alpha_{2}\right\}, \{\beta_{1}, \beta_{2}\right\}\right)_{\mathcal{G} \oplus \mathcal{G}}:=\left(\alpha_{1}, \beta_{1}\right)-\left(\alpha_{2}, \beta_{2}\right)
\end{equation}
This isomorphism $\Phi_\lambda$ is used to go from $\mathcal{E}$ to $\mathcal{E}_\lambda$ where, $\mathcal{E}: \mathcal{D} \rightarrow \mathcal{D}$ and $\mathcal{E}_{\lambda}: \mathcal{G} \oplus \mathcal{G} \rightarrow \mathcal{G} \oplus \mathcal{G}$ such that,
\begin{equation}\label{E cond}
\mathcal{E}_{\lambda} \circ \Phi_{\lambda}=\Phi_{\lambda} \circ \mathcal{E}~~~~~~~~~~\text{where,}~~~~~~\mathcal{E}\left(x_{1} \dot{+} x_{2}\right)=\left(x_{2} \dot{+} x_{1}\right)
\end{equation}
The real linear map $\mathcal{E}_\lambda:\mathcal{G}\oplus\mathcal{G}\rightarrow\mathcal{G}\oplus\mathcal{G}$ which depends on $\lambda$, is constructed by demanding it is idempotent and self-adjoint under the inner-product of the algebra $\mathcal{G}\oplus\mathcal{G}$ as described in (\ref{cond for E map}). A generic form in terms of arbitrary matrices acting on $X,Y\in \mathcal{G}$ is assumed for $\mathcal{E}_\lambda$ and the constraints (\ref{cond for E map}) are solved for these matrices to obtain the mapping explicitly (detailed derivation can be found in Appendix \ref{C}),
\begin{equation}\label{E map}
\begin{aligned}
\mathcal{E}_\lambda(\{X,Y\})=&\{\big[\left(1-\lambda \lambda^{T}\right)^{-1}+\left(\lambda^{-T} \lambda^{-1}-1\right)^{-1}\big]X-2\left(\lambda^{-1}-\lambda^{T}\right)^{-1}Y,\\
&2\left(\lambda^{-T}-\lambda\right)^{-1} X-\big[\left(1-\lambda^{T} \lambda\right)^{-1}+\left(\lambda^{-1} \lambda^{-T}-1\right)^{-1}\big]Y\} 
\end{aligned}
\end{equation}
here $X,Y\in \mathcal{G}$. The above map $\mathcal{E}_\lambda$ is related to the map $\mathcal{E}$ (\ref{E cond}) via the following isomorphism $\Phi_\lambda$,
\begin{equation}\label{phi map}
\begin{aligned}
\Phi_{\lambda}\left(x_{1} \dot{+} x_{2}\right)=\{x_{1}&+\left(1-\lambda \lambda^{T}\right)^{-1}\left[1+\lambda \lambda^{T}-2 \lambda\right] x_{2},\\
&x_{1}-\left(1-\lambda^{T} \lambda\right)^{-1}\left[1+\lambda^{T} \lambda-2 \lambda^{T}\right] x_{2}\}
\end{aligned}
\end{equation}
This isomorphism as desired respects the inner product of the double $\mathcal{D}$ as well as the inner product for the direct sum $\mathcal{G}\oplus\mathcal{G}$.
We aim to compute the action for this $\mathcal{E}-$model given by ($\ref{action DFT}$) which requires a projection operator $P_{f}(\mathcal{E}_\lambda)$ such that,
\begin{equation}\label{3.7}
\begin{aligned}
\operatorname{Im} P_{f}\left(\mathcal{E}_{\lambda}\right) &=\mathcal{G}^{\delta}=(\alpha, \alpha) \\
\operatorname{Ker} P_{f}\left(\mathcal{E}_{\lambda}\right) &=\left(\mathbb{I}+D_{f-1} \mathcal{E}_{\lambda} D_{f}\right) \mathcal{G}^{\delta}
\end{aligned}
\end{equation}
where, $\tilde{\mathcal{H}}=\mathcal{G}^\delta$ is the Lie algebra for the diagonal subgroup $G_{\text{diag}}$ of $G\times G$ which has the elements of the type $\{\alpha,\alpha\}$ where $\alpha\in\mathcal{G}$. The Projection operator is uniquely determined by using the constraints given above in (\ref{3.7}) to get the generalized $\lambda-$model action and is given by,
\begin{equation}\label{projector}
\begin{aligned}
P_{\{g, e\}}\left(\mathcal{E}_{\lambda}\right)(\alpha, \beta)=\bigg((1-&D^{-1} \lambda^{-T})^{-1} \alpha+\left(1-\lambda^{T} D\right)^{-1} \beta,\\
&\left(1-D^{-1} \lambda^{-T}\right)^{-1} \alpha+\left(1-\lambda^{T} D\right)^{-1} \beta\bigg)
\end{aligned}
\end{equation}
Using this projection operator in the action from (\ref{action DFT}) for $f=\{g,e\}\in G\times G/G_{\text{diag}}$ where $g\in {G}$ and $e$ is the identity,
\begin{equation}
\begin{aligned}
S_{\mathbb{D} / \tilde{H}} &= S_{W Z W}[f]-\frac{1}{\pi} \int d \sigma d \tau\big(\mathcal{P}_f(\mathcal{E}_\lambda)\left(f^{-1} \partial_{+} f\right), f^{-1} \partial_{-} f\big)_{\mathcal{G}\oplus\mathcal{G}}\\
\end{aligned}
\end{equation}we reproduce the action for the generalized $\lambda-$deformed model given by (\ref{lambda action}),
\begin{equation}
\begin{aligned}
&S_{\lambda}(g)=S_{\text{WZW}}(g) + \int d \xi^{+} d \xi^{-}\left(\partial_{+} g g^{-1},({\lambda}^{-1}- \mathrm{Ad}_{g^{-1}}\right)^{-1} g^{-1} \partial_{-} g)
\end{aligned}
\end{equation}
This shows how we can embed the generalized $\lambda$-model into a $\mathcal{E}-$model by constructing an appropriate $\mathcal{E}-$map. The $\mathcal{E}-$map (\ref{E map}) and the isomorphism $\Phi_\lambda$ (\ref{phi map}) obtained here are consistent with the results obtained in \cite{klimvcik2015eta} as they boil down to the following corresponding mappings for the case when $\lambda$ is assumed to be a number (denoted by $\tilde{\lambda}$ to distinguish from the case when $\lambda$ is a matrix),
\begin{equation}
\begin{aligned}
\mathcal{E}_{\tilde{\lambda}}(\{X,Y\})=&\frac{1}{2}\bigg(\frac{1-\tilde{\lambda}}{1+\tilde{\lambda}}+\frac{1+\tilde{\lambda}}{1-\tilde{\lambda}}\bigg)\{X,-Y\}+\frac{1}{2}\bigg(\frac{1-\tilde{\lambda}}{1+\tilde{\lambda}}-\frac{1+\tilde{\lambda}}{1-\tilde{\lambda}}\bigg)\{Y,-X\}\\
&\Phi_{\tilde{\lambda}}\left(x_{1} \dot{+} x_{2}\right)=\left\{x_{1}+\frac{1-\tilde{\lambda}}{1+\tilde{\lambda}} x_{2}, x_{1}-\frac{1-\tilde{\lambda}}{1+\tilde{\lambda}} x_{2}\right\}
\end{aligned}
\end{equation}
In order to find the DFT description of the generalized $\lambda-$model we further use the derived $\mathcal{E}-$map to obtain the generalized metric $\mathcal{H}_{AB}$ in the following section.
\subsection{Generalized metric $\mathcal{H}_{AB}$ and $\widehat{\mathcal{H}}_{\hat{I}\hat{J}}$}\label{3.2}
The generalized metric $\mathcal{
H}_{AB}$ along with the generalized frame fields will be used to obtain the coordinate dependent generalized metric $\widehat{\mathcal{H}}_{\hat{I}\hat{J}}(x)$ which defines the DFT corresponding to the worldsheet $\mathcal{E}-$model derived earlier. We use the derived $\mathcal{E}-$map (\ref{E map}) and the constraint (\ref{cond for E and H}) that parametrizes $\mathcal{E}_A{}^B$ in terms of the generalized metric to determine $\mathcal{H}_{AB}$ as follows (detailed derivation can be found in the appendix \ref{C}),
\begin{equation}\label{H unrotated}
\mathcal{H}_{A B}=\frac{1}{2}\left(\begin{array}{ll}
\left(A_{1}-A_{2}+B_{1}-B_{2}\right)\kappa^{-1} & \left(A_{1}+A_{2}+B_{1}+B_{2}\right)\kappa^{-1} \\
\left(A_{1}-A_{2}-B_{1}+B_{2}\right)\kappa & \left(A_{1}+A_{2}-B_{1}-B_{2}\right)\kappa
\end{array}\right)
\end{equation}
where, 
\begin{equation}\label{A_i,B_i}
\begin{aligned}
&A_1=\big[(1-\lambda\lambda^T)^{-1}+(\lambda^{-T}\lambda^{-1}-1)^{-1}\big]~~~~~~~~~A_2=-2(\lambda^{-1}-\lambda^T)^{-1}\\
&B_2=-\big[(1-\lambda^T\lambda)^{-1}+(\lambda^{-1}\lambda^{-T}-1)^{-1}\big]~~~~~~~B_1=2(\lambda^{-T}-\lambda)^{-1}
\end{aligned}    
\end{equation}
where, $\eta_{AB}$ is given by (\ref{eta}). It has been established in \cite{demulder2019doubled} that for each group $\mathbb{D}$ with a non-degenerate, ad-invariant bilinear form $\eta$ and a maximally isotropic subgroup $\tilde{H}$, a generalized frame field $\widehat{E}_{A}{}^{\hat{I}}\in O(D,D)$ that depends only on the coordinates $x^i$ can be defined on the symmetric space $\mathbb{D}/\tilde{H}$. 
\par Here, such a generalized frame field can be defined on the space $G\times G/G_{\text{diag}}$ which is parametrized by $f\in G\times G/G_{\text{diag}}$. \cite{demulder2019doubled} uses the parametrization $f=\{\bar{g},\bar{g}^{-1}\}$ for $\bar{g}\in G$ which defers to the convention $f=\{g,e\}$ followed here \cite{klimvcik2015eta} in defining the $\mathcal{E}-$model. In order to match the conventions, an identification $\bar{g}^2=g\in G$ is employed. The generalized frame field $\widehat{E}_{A}{}^{\hat{I}}$ is ultimately given by \cite{demulder2019doubled},
\begin{equation}
\widehat{E}_{A}^{ I}=\frac{1}{2}\tensor{\left(\begin{array}{cc}
\frac{1}{\sqrt{2}} (1+D) & \sqrt{2} \kappa^{-1}(1-D) \\
\frac{1}{\sqrt{2}}(1-D) \kappa & \sqrt{2}(1+D)
\end{array}\right)}{_{A}^{B}}\tensor{\left(\begin{array}{cc}
e & 0 \\
0 & e^{-T}
\end{array}\right)}{_{B}^{\hat{I}}}
\end{equation}
where, $e$ is the left-invariant form and $D=\operatorname{Ad}_g$ is the adjoint action of $G$. The adjoint action $D$ is orthogonal, $D^T=D^{-1}$ and hence commutes with $\kappa$ and $\kappa^{-1}$. The generalized frame field can be decomposed in terms of $\rho,b,\beta$ as
    \begin{equation}\label{frame field}
\widehat{E}_{A}^{} \hat{I}=\left(\begin{array}{cc}
\rho^{-T} & 0 \\
0 & \rho
\end{array}\right)\left(\begin{array}{ll}
1 & 0 \\
b & 1
\end{array}\right)\left(\begin{array}{ll}
1 & \beta \\
0 & 1
\end{array}\right)\left(\begin{array}{cc}
e & 0 \\
0 & e^{-T}
\end{array}\right)
\end{equation}
where,
\begin{equation}
\rho^{-T}=\frac{1}{2 \sqrt{2}}\left(1+D\right), \quad b=\frac{1}{8}\left(D^{-1}-D\right) \kappa \quad \text { and } \quad \beta=2 \kappa^{-1} \frac{1-D}{1+D}
\end{equation}
This decomposition will be used for comparing with the generalized frame fields used in \cite{borsato2021supergravity} and in the following section. This decomposition in terms of the parameters $\rho, b$ and $\beta$ is the most general parametrization for $O(D,D)$ where, $\beta^{ij}$ and $b_{ij}$ are anti-symmetric and the matrix $\rho\in GL(D)$.
\par We can now gather the generalized frame field as stated above (\ref{frame field}) and the generalized metric $\mathcal{H}_{AB}$ obtained in (\ref{H unrotated}), in the following relation to determine the DFT coordinate dependent metric $\widehat{\mathcal{H}}^{\hat{I}\hat{J}}(x)$
\begin{equation}\label{H H relation}
\widehat{\mathcal{H}}^{\hat{I}\hat{J}}(x)=\widehat{E}_{A}^{}{}^{\hat{I}}\mathcal{H}^{AB}\widehat{E}_{B}^{}{}^{\hat{J}}
\end{equation}
\par To visualize the embedding of the non-linear $\sigma-$model we can compare the known metric $\widehat{\mathcal{H}}^{\hat{I}\hat{J}}(x)$ obtained from (\ref{H H relation}) to the known format of the metric to read off the metric $G$ and $B$-field of the $\sigma-$model,
\begin{equation}\label{Hij}
\widehat{\mathcal{H}}_{\hat{I}\hat{J}}(x)=\begin{pmatrix}
G^{-1}&-G^{-1}B\\
BG^{-1}&G-BG^{-1}B
\end{pmatrix}_{\hat{I}\hat{J}}
\end{equation}
This gives the following metric and B-field which precisely match with the results known for a generalized $\lambda-$deformed model in \cite{sfetsos2015generalised, demulder2015integrable,20141},
\begin{equation}\label{metric result}
ds^2=\frac{1}{2}\big[(\hat{O}_{g^{-1}}+\hat{O}_g-1)\kappa\big]_{ab}e^a\otimes e^b
\end{equation}
\begin{equation}\label{B-field result usual}
B=B_{WZW}+\frac{1}{4}\big[\hat{O}_{g^{-1}}-\hat{O}_g\big]_{ab}e^a\wedge e^b
\end{equation}
where, 
\begin{equation}
\hat{O}_g=(1-\lambda D)^{-1}~~~~~~~~~~~~~\hat{O}_{g^{-1}}=(1-D^{-1}\lambda^T)^{-1}    
\end{equation}
The above expressions (\ref{Hij}), (\ref{metric result}) and (\ref{B-field result usual}) provide the supergravity embedding of the generalized $\lambda-$deformed model in the doubled formalism of a DFT. Here, we have only focused on finding the metric and the B-field and evaluation of the RR fields require separate investigation. 
\section{Asymmetrical generalized $\lambda-$deformation}\label{4}
The generalized $\lambda-$deformed model on a symmetric space $\mathbb{D}/\tilde{H}=G\times G/G_{\text{diag}}$ is extended to incorporate the possibility of deforming the left-right asymmetrically gauged WZW model \cite{2019,2021}. This modified model is constructed by gauging the following,
\begin{equation}
g\rightarrow h^{-1}gh^\prime  ~~~~~~~~~~~~~\hat{g}\rightarrow h^{-1}\hat{g}   
\end{equation}
instead of,
\begin{equation}
g\rightarrow h^{-1}gh~~~~~~~~~~~~~~~\hat{g}\rightarrow \hat{g}h   
\end{equation}
where, the WZW and PCM model are defined as (\ref{WZW}) and (\ref{PCM}) respectively, along with $h=e^{G^a{t}_a}\in G$ and $h^\prime=e^{G^a{{t}}_a^\prime}\in G$ such that ${{t}^\prime}_a=Wt_a$. The $W$ is a constant outer automorphism of the algebra which preserves $\kappa$. Upon gauge fixing by setting $\hat{g}=\mathbb{I}$, the asymmetrical $\lambda-$model is given by \cite{2019},
\begin{equation}\label{asym action}
S_\lambda(g,W)=S_{\text{WZW}}  +\frac{1}{\pi}\int d\sigma  d\tau \langle\partial_+gg^{-1},(\lambda^{-1}-D_gW)^{-1}\partial_-gg^{-1}\rangle  
\end{equation}
This deforms the left-right asymmetrically gauged $\mathbb{D}/\tilde{H}_{\text{AS}}$ WZW model instead of vectorially gauged $\mathbb{D}/\tilde{H}$ WZW model. As followed in the previous section, we aim here to give a doubled framework in terms of an $\mathcal{E}-$model and DFT for the asymmetrical generalized $\lambda-$deformed model described by (\ref{asym action}). It has been suggested in \cite{borsato2021supergravity} that the parametrization and hence the generalized frame field which describes the DFT for the symmetrically gauged $\lambda$-model ($\lambda$ as a number) can also be used to describe the asymmetrical extension of the model. Hence, we shall use the results obtained in the previous section for the asymmetrically gauged generalized model.
\par Following the same procedure, we develop the $\mathcal{E}-$model to find the generalized metric $\mathcal{H}_{AB}$. The action (\ref{asym action}) is obtained by first using the parametrization $f=\{\bar{g},\bar{g}^{-1}\}$ with corresponding left-right form and $\bar{g}^2=\tilde{g}\in G$ is set to respect the conventions. Another parametrization is considered to incorporate the $W$ such that $f=\{g,g^{-1}\}$ such that the adjoint action is non-trivially related to the parametrization as, $\operatorname{Ad}_{\tilde{g}}=\operatorname{Ad}_{g}.W$. \cite{borsato2021supergravity} visualizes the working of $W$ by considering a constant $W\in \exp{\mathcal{G}}$ such that $\tilde{g}=gw$ and defines $W(t_i)=wt_iw^{-1}$ indicating that $W$ is orthogonal. The $\mathcal{E}-$map (\ref{E map}) and the projection operator (\ref{projector}) written for the element $\tilde{g}$ are used to compute the action. To address the embedding and DFT, we require the generalized frame field, that by using the parametrization (\ref{frame field})\footnote{There are slight differences in how the decomposition is used in \cite{borsato2021supergravity} as compared to the one used in the previous section and the change has been incorporated in the ($\rho,\beta, b$) given here. The decomposition is as follows ($e=g^{-1}dg$ is replaced by $v=dgg^{-1}$ here),
\begin{equation}
E\sim\begin{pmatrix}
1&\beta\\
0&1
\end{pmatrix}\begin{pmatrix}
1&0\\
b&1
\end{pmatrix}
\begin{pmatrix}
\rho^T&0\\
0&\rho^{-1}
\end{pmatrix}
\begin{pmatrix}
v^T&0\\
0&v
\end{pmatrix}
\end{equation}} is redefined in terms of $\rho,\beta$ and $b$ to be \cite{borsato2021supergravity}, 
\begin{equation}
\begin{gathered}
\rho^{-1}=\frac{1}{\sqrt{2}}\left(1+\operatorname{Ad}_{g} W\right) v^{-1}, \quad \beta=-\kappa^{-1} \frac{1-\operatorname{Ad}_{g} W}{1+\operatorname{Ad}_{g} W} \\
b=\frac{1}{4}\left(\operatorname{Ad}_{g} W-W^{-1} \mathrm{Ad}_{g}^{-1}\right) \kappa
\end{gathered}
\end{equation}
Since the same $\mathcal{E}-$map from (\ref{E map}) is used, we work with the same generalized metric $\mathcal{H}_{AB}$ as given in (\ref{H unrotated}), and the relation (\ref{cond for E and H}) to find the coordinate dependent generalized metric $\widehat{\mathcal{H}}_{\hat{I}\hat{J}}$ to identify the DFT for the asymmetrical model. The following information of the embedded metric $G$ and $B-$field is read off from the known format of $\widehat{\mathcal{H}}_{\hat{I}\hat{J}}$ (\ref{Hij}),
\begin{equation}
    G=\frac{1}{2}\big[(\hat{O}^w_{g^{-1}}+\hat{O}^w_g-1)\kappa\big]_{ab}e^a\otimes e^b
\end{equation}
\begin{equation}\label{B-field result}
B=B_{WZW}+\frac{1}{2}\big[\hat{O}^w_{g^{-1}}-\hat{O}^w_g\big]_{ab}e^a\wedge e^b
\end{equation}
where, 
\begin{equation}
\hat{O}^w_g=(1-\lambda DW)^{-1}~~~~~~~~~~~~~\hat{O}^w_{g^{-1}}=(1-W^{-1}D^{-1}\lambda^T)^{-1}    
\end{equation}
The above match the previously obtained results in (\ref{metric result}) and (\ref{B-field result usual}) for the symmetric case (considering the change in conventions) when $W$ is set to 1. Therefore this extends the doubled framework of DFT for the asymmetrical generalized $\lambda-$deformed model as well. 
\par Given a DFT formulation in terms of the generalized frame field and the underlying Drifeld algebra, article \cite{borsato2021supergravity} classified various parametrizations into orbits identified by the non-zero generalized flux components. The algebraic structures involved in this classification were expressed in terms of frame fields and Drinfeld doubles, and these ingredients do not depend on the deformation parameters. Therefore, the classification of \cite{borsato2021supergravity} works for the generalized $\lambda$-deformation as well. For completeness, we briefly review this construction here.  

To classify the orbits, the authors of \cite{borsato2021supergravity} began with writing the commutation relations (\ref{commutation relation}) in more explicit form as\footnote{Recall that the generators of the Lie algebra $\mathcal{D}$ are given by $\mathbb{T}_A=(\tilde{T}^a,T_a)$, where $\{\tilde{T}^a\}$ are the generators of $\tilde{{H}}$, the maximally isotropic subgroup of $\mathbb{D}$, and $\{{T}_a\}$ span the complementary $\mathbb{D}/\tilde{H}$.}
\begin{equation}
[T_a,T_b]=F_{ab}{}^cT_c+H_{abc}\tilde{T}^c\quad\quad [T_a,\tilde{T}^b]=Q_a{}^{bc}T_c-F_{ac}{}^b\tilde{T}^c\quad\quad[\tilde{T}^a,\tilde{T}^b]=Q_c{}^{ab}\tilde{T}^c+R^{abc}T_c    
\end{equation}
The orbits were labelled by the non-vanishing fluxes out of $F,Q,H,R$. A particular orbit covered all the different parameterizations (generalized frame fields) that gave the same generalized fluxes.
\par For the generalized $\lambda-$deformation model, we considered the double $\mathcal{G}\oplus\mathcal{G}$ and took the maximally isotropic subgroup to be the diagonal $\mathcal{G}_{\text{diag}}$. The fluxes for the undeformed theory are given by \cite{demulder2019doubled}
\begin{equation}\label{fluxes}
H_{abc}=\frac{1}{\sqrt{2}} f_{ab}{}^d \kappa_{dc}, \quad Q_{a}{}^{b c}=\frac{1}{\sqrt{2}} \kappa_{a d} \kappa^{b e} \kappa^{c g} f_{e g}{}^{d}, \quad F_{ab}{ }^{c}=0, \quad R^{abc}=0,
\end{equation}
where $f_{ab}{}^c$ are the structure constants of $\mathcal{G}$. %The above structure is the result of the fact that the complementary $G_{\text{anti-diag}}$ does not form a subgroup and is merely a symmetric space.
Since the classification of orbits depends only on the generalized frame fields (parametrization) and the Drinfeld double structure, but not on the deformation parameters, the authors of \cite{borsato2021supergravity} concluded that the (asymmetrical) $\lambda-$deformation model falls into the $(H,Q)-$orbit, which has only $H$ and $Q$ fluxes. The generalized $\lambda$-deformation has the same underlying algebraic structure (\ref{fluxes}), therefore, it also falls into the $(H,Q)-$orbit, for both symmetric and asymmetric deformations.
\section{Conclusion and outlook}\label{5}
In this work, we extended the doubled formalism of $\mathcal{E}-$model and DFT for the generalized $\lambda-$deformation. Our results (\ref{H unrotated})-(\ref{A_i,B_i}) and (\ref{metric result})-(\ref{B-field result usual}) are in agreement with the supergravity equations as the section condition was satisfied by ensuring the manifold was created by cosetting a maximally isotropic subgroup $\tilde{H}$ of $\mathbb{D}$. We also commented on the asymmetrical generalized $\lambda-$deformation and its embedding into DFT. It was worth noting to see that the parametrization used for usual the $\lambda-$model also resulted in the DFT for the generalized model. Correspondingly, the results obtained in this paper boil down to the outcomes of the usual model when $\lambda$ is assumed to be a simple parameter instead of a matrix.
\par Here, we only focused on obtaining the metric and the  Kalb-Ramond field from the DFT framework. However, this can be further continued to a complicated calculation of finding the dilaton and the RR fields as performed in \cite{demulder2019doubled}. T-dualities in non-isometric directions are also made possible in the framework of DFT by having dependence on the dual coordinates. Since the doubled formalism highlights hidden symmetries, it would be interesting to use the methods developed in this article to identify such non-local symmetries for the generalized $\lambda$-deformations of specific manifolds like $\operatorname{AdS}_p\times \operatorname{S}^q$ \cite{2016,2018,20161,2004}.

\subsection*{Acknowledgements} 
\par I would like to thank my advisor Oleg Lunin for initiating this project and immense support throughout this work. This work was supported in part by the DOE grants DE-SC0017962 and DE-SC0015535.

\appendix
\section*{Appendix}
\addcontentsline{toc}{section}{Appendices}
\renewcommand{\thesubsection}{\Alph{subsection}}
\numberwithin{equation}{subsection}
\subsection{Conventions}\label{A}
We follow the notations used in \cite{demulder2019doubled,20131} throughout this paper. The indices $A,B,\ldots=1,2,\ldots,2D$ are the flat $O(1,D-1)\times O(D-1,1)$ indices and $I,J,\hat{I},\hat{J},\ldots=1,2,\ldots,2D$ are the algebra indices where the hatted indices $\hat{I},\hat{J},\ldots$ indicate their dependence strictly on the coordinates $x^i$ and not on the dual coordinates $\tilde{x}_{\tilde{i}}$. The indices $M,N,\ldots$ are the curved $O(D,D)$ indices. 
\par The real Lie algebra $\mathcal{D}$ has the corresponding group $\mathbb{D}$ with generators $\{\mathbb{T}_A\}$ that obey the commutation relations as given in (\ref{commutation relation}). The doubled coordinates here are denoted by $X^I=(x^i,\tilde{x}_{\tilde{i}})$. The maximally isotropic subgroup for $\mathbb{D}$ is denoted by $\tilde{H}$ with the corresponding algebra $\tilde{\mathcal{H}}$. The generators for $\mathcal{H}$ are given by $\{\tilde{T}^a\}$ with structure constants $\tilde{F}^{ab}{}_c$ and the space is spanned by the dual coordinates $\tilde{x}_{\tilde{i}}$.
\par The generalized $\lambda-$model is defined on the group $G$ with corresponding algebra denoted by $\mathcal{G}$. The generators of the algebra are $\{t_a\}$ with inner product $(t_a,t_b)=\kappa_{ab}$ and structure constants $f_{ab}{}^c$. The maximally isotropic subgroup for the double $G \times G$ is given by the diagonal subgroup $G_{\text{diag}}$ generated by $x\to\{x,x\}/\sqrt{2}$. The complementary isotropic symmetric space is the anti-diagonal group $G_{\text{anti-diag}}$. The left ($e$) and right ($v$) invariant Maurer-Cartan forms in terms of the element $g\in G$ for algebra $\mathcal{G}$ obey the following,
\begin{equation}
\begin{aligned}
&g^{-1}dg=e^at_a=\tensor{e}{^a_i}dx^it_a~~~~~~~~~~~~~~~~~~dgg^{-1}=v^at_a=\tensor{v}{^a_i}dx^it_a\\
&\operatorname{Ad}_gh=Dh=D_gh=ghg^{-1}~~~~~~~~~~~\operatorname{Ad}_{g^{-1}}h=D^{-1}h=D_{g^{-1}}h=g^{-1}hg\\
&v=De~~~~~~~~~D^T=D^{-1}~~~~~~~~~~~~~~~~\kappa D=D\kappa~~~~~~~~~~\kappa D^T=D^T \kappa
\end{aligned}
\end{equation}
In literature, the generalized $\lambda-$deformation is sometimes also identified by the following action \cite{sfetsos2015generalised},
\begin{equation}\label{alternate action}
\begin{aligned}
&S_{\lambda}(g)=\frac{1}{2} \int d \xi^{+} d \xi^{-}\left(g^{-1} \partial_{+} g, g^{-1} \partial_{-} g\right)+\frac{1}{12} \int d^{-1}\left(d g g^{-1},\left[d g g^{-1}, d g g^{-1}\right]\right) \\
&\quad+ \int d \xi^{+} d \xi^{-}\left(g^{-1}\partial_+g,(\lambda^{-1}-\operatorname{Ad}_g)^{-1}\partial_-gg^{-1}\right)
\end{aligned}
\end{equation}
The actions in (\ref{alternate action}) and (\ref{lambda action}) can be both used as they are related by the element inversion given below. In order to respect the conventions followed throughout the paper, we shall consider the action to be given by (\ref{lambda action}) as per \cite{2014}, however, the following transformation can be used to get the results aiming for the action in (\ref{alternate action}).
\begin{equation}
g\rightarrow g^{-1}:~~L^a\leftrightarrow -R^a~~~~D\leftrightarrow D^T~~~~~~~e^a\leftrightarrow -\tilde{e}^a
\end{equation}
\subsection{Embedding bi-YB $\sigma$-model into generalized $\lambda$-model}\label{app eta}
In this appendix, we summarize the connection between the bi-Yang-Baxter (bi-YB) $\sigma$-model and the generalized $\lambda$-deformation \cite{1606.03016, sfetsos2015generalised}. The bi-YB model was introduced in \cite{1402.2105} and the action reads,
\begin{equation}
S_{\text{bi-YB}}=\frac{1}{2\pi t}\int d^2\sigma(g^{-1}\partial_+g,(1-\eta\mathcal{R}-\rho\mathcal{R}_g)^{-1}g^{-1}\partial_-g). 
\end{equation}
Here, $\eta$ and $\rho$ are the two deformation parameters and $\mathcal{R}_g=D_{g^{-1}}\mathcal{R}D_{g}$. The anti-symmetric matrix $\mathcal{R}$ is a solution of the modified YB equation,
\begin{equation}
[\mathcal{R}A,\mathcal{R}B]-\mathcal{R}([\mathcal{R}A,B]+[A,\mathcal{R}B])=-c^2[A,B]\quad\quad \text{for,~}A,B\in \mathcal{G},\quad c\in \mathbb{C}    
\end{equation}
Following earlier studies of relationship between a single parameter YB $\sigma$-model and $\lambda$-deformation \cite{klimvcik2015eta}, article
\cite{1606.03016} related the generalized $\lambda$-deformation of a YB $\sigma$-model to a bi-YB $\sigma$-model by a Poisson-Lie T-duality and analytic continuation\footnote{For the $SU(2)$ group and $SU(2)/U(1)$ coset, this was done earlier in \cite{sfetsos2015generalised}.}.
Specifically starting with the bi-YB model and rewriting its PL T-dual as
\begin{equation}
S_{\text{PL T-dual}}(p)=-iS_{\text{WZW}}(p^2)-ik\int d^2\sigma\Bigg(\bigg(\frac{1+i\eta+\rho\mathcal{R}}{1-i\eta+\rho\mathcal{R}}-D_{p^2}\bigg)^{-1}\partial_+(p^2)p^{-2},p^{-2}\partial_-(p^2)\Bigg), 
\end{equation}
the author of \cite{1606.03016} demonstrated that this is an analytically continued generalized $\lambda$-model (\ref{lambda model})\footnote{The explicit analytic continuation between $p$ and $g$ can be found in \cite{1606.03016}. },
\begin{equation}\label{pl lambda}
S_{\text{gen.}\lambda}(g)=S_{WZW}(g)+k\int d^2\sigma((\lambda^{-1}-D_{g})^{-1}\partial_+gg^{-1},g^{-1}\partial_-g)
\end{equation}
with 
\begin{equation}\label{old one}
\lambda^{-1}=\frac{1+i\eta+\rho\mathcal{R}}{1-i\eta+\rho\mathcal{R}}
\end{equation}
On the other hand, starting with the YB-model
\begin{equation}
S_{YB}=\frac{1}{2\pi\tilde{t}}\int d^2\sigma(g^{-1}\partial_+g,(1-\tilde{\eta}\mathcal{R})^{-1}g^{-1}\partial_-g),
\end{equation}
and deforming it according to (\ref{lambda model}) with
\begin{equation}\label{F choice}
\hat{F}=\frac{1}{\tilde{t}}(\mathbb{I}-\tilde{\eta}\mathcal{R})^{-1}
\end{equation}
one finds (\ref{pl lambda}) with 
\begin{equation}\label{new one}
\lambda^{-1}=\frac{1}{k\tilde{t}}(\mathbb{I}-\tilde{\eta}\mathcal{R})^{-1}+\mathbb{I}    
\end{equation}
the deformation parameters (\ref{old one}) and (\ref{new one}) are related by an analytic continuation,
\begin{equation}
\rho=-\frac{2k\tilde{t}\tilde{\eta}}{2k\tilde{t}+1},\quad\eta=-\frac{i}{2k\tilde{t}+1}
\end{equation}
\par To summarize, in this appendix we have incorporated generalized $\eta$-model into the framework of the generalized $\lambda$-deformation. In section \ref{3}, we embed the latter into the DFT formalism, thus providing such embedding for the integrable generalized $\eta$-model as well. 

\subsection{Background on $\mathcal{E}-$model}\label{B}
In this appendix, we review some of the basics of the $\mathcal{E}-$model. For a real linear algebra $\mathcal{D}$ as defined above, an infinite dimensional Poisson manifold $P_{\mathcal{D}}$ is constructed, parametrized by co-ordinates $j^A(\sigma)$ such that the Poisson bracket is given by,
\begin{equation}\label{PB}
\left\{j^{A}(\sigma), j^{B}\left(\sigma^{\prime}\right)\right\}_{\text{P.B.}}=\mathbb{F}^{A B}{}_C j^{C}(\sigma) \delta\left(\sigma-\sigma^{\prime}\right)+\eta^{A B} \partial_{\sigma} \delta\left(\sigma-\sigma^{\prime}\right)
\end{equation}
here, $\mathbb{F}^{AB}{}_C$ are the structure constants of $\mathcal{D}$, indicating that the $P_{\mathcal{D}}$ plays the role of the current algebra for $\mathcal{D}$. The map $j=j^A\mathbb{T}_A$ takes values in $\mathcal{D}$. A Hamiltonian $H_{\mathcal{E}}$ in $j^A(\sigma)$ is obtained using the $\mathcal{E}-$map
\begin{equation}
\text { H}_\mathcal{E}:=\frac{1}{2} \oint d \sigma\big( j(\sigma), \mathcal{E}(j(\sigma)) \big)_\mathcal{D}
\end{equation}
The $\mathcal{E}-$model is defined as the dynamical system on the phase space given by $P_{\mathcal{D}}$ and the Hamiltonian $H_{\mathcal{E}}$, giving a doubled description for a worldsheet theory. The Lie algebra $\mathcal{D}$ is assumed to have a linear one-parameter family of the structure constants such that,
\begin{equation}
\mathbb{F}^{A B}{}_{C}=\mathbb{F}_{0}^{A B}{}_C+\varepsilon \mathbb{F}_{1}^{A B}{}_C, \quad \varepsilon \in \mathbb{R}
\end{equation}
with the corresponding Poisson brackets given by,
\begin{equation}
\left\{j^{A}(\sigma), j^{B}\left(\sigma^{\prime}\right)\right\}_{\text{P.B.}}=\left\{j^{A}(\sigma), j^{B}\left(\sigma^{\prime}\right)\right\}_{0}+\varepsilon\left\{j^{A}(\sigma), j^{B}\left(\sigma^{\prime}\right)\right\}_{1}
\end{equation}
The parameter $\epsilon$ holds quite a significance in the context of $\mathcal{E}-$models. \cite{klimvcik2015eta} describes how the $\epsilon$ acts as an umbrella for various $\sigma-$models that are described by the $\mathcal{E}-$models. The principal chiral model is embedded into the formalism of $\mathcal{E}$-model for the case when $\epsilon=0$, while for negative $\epsilon$ values, the $\eta$-deformed model is addressed. $\epsilon$ can also be thought of as the parameter involved like $\lambda$ or $\eta$ when dealing with the corresponding model. \cite{klimvcik2015eta} has described $\epsilon$ in terms of $\tilde{\lambda}$ for the $\lambda-$deformed model as $\epsilon=\frac{1-\tilde{\lambda}}{1+\tilde{\lambda}}$ and it was observed that in this doubled formalism $\epsilon$ could see more values of $\tilde{\lambda}$ as $\tilde{\lambda}$ could now range between $\{-1,1\}$ instead of $\{0,1\}$ for a positive $\epsilon$. This emphasizes one of the benefits of working in $\mathcal{E}-$models and DFTs.
\subsection{Deriving $\mathcal{E}_\lambda-$map and isomorphism $\Phi_\lambda$}\label{C}
In section \ref{3}, we work in the double algebra $\mathcal{G}\oplus\mathcal{G}$ instead of the algebra $\mathcal{D}$ and use the isomorphism $\Phi_\lambda$ to map the two algebras. Therefore, we look at the derivation for the real linear map $\mathcal{E}_\lambda:\mathcal{G}\oplus\mathcal{G}\rightarrow\mathcal{G}\oplus\mathcal{G}$ here. We assume a general form for $\mathcal{E}_\lambda-$map as following for some arbitrary matrices $A_i$ and $B_i$,
\begin{equation}\label{E skeleton}
\mathcal{E}_\lambda(\{X,Y\})=\{A_1X+A_2Y,B_1X+B_2Y\}~~~~~~~~~~~~~\text{where,}~~X,Y\in \mathcal{G}
\end{equation}
The $\mathcal{E}_\lambda-$map is required to be idempotent and self-adjoint under $(.,.)_{\mathcal{G}\oplus\mathcal{G}}$ (\ref{GG inner product}) which can be written as following for $x,y\in \mathcal{G}\oplus\mathcal{G}$,
\begin{equation}
(\mathcal{E}_\lambda x,y)_{\mathcal{G}\oplus\mathcal{G}}=(x,\mathcal{E}_\lambda y)_{\mathcal{G}\oplus\mathcal{G}}~~~~~~~~~~~~~~~~~~~~\mathcal{E}_\lambda x^2=x    
\end{equation}
We use the general form (\ref{E skeleton}) to rewrite the above equations as constraints on the matrices $A_i$ and $B_i$,
\begin{equation}\label{skeleton condition}
\begin{aligned}
& A_1^T=A_1~~~~~~~~B_2^T=B_2~~~~~~~~A_2=-B^T_1\\
& A_1A_2+A_2B_2=0~~~~~~~~~B_1A_1+B_2B_1=0\\
& B_1A_2+B_2B_2=1~~~~~~~~~A_1A_1+A_2B_1=1
\end{aligned}    
\end{equation}
\cite{demulder2019doubled} makes a particular choice for the isomorphism $\Phi_\lambda$\footnote{The isomorphism $\Phi_\lambda$ and $\mathcal{E}_\lambda$-map here are denoted by $\Phi_\epsilon$ and $\mathcal{E}_\epsilon$ respectively in \cite{demulder2019doubled}} to develop the $\mathcal{E}-$model for the $\lambda-$model. Here, we derive the isomorphism $\Phi_\lambda$ using the $\mathcal{E}_\lambda$-map, hence we make the following choice for matrix $B_1$ instead,
\begin{equation}\label{B_1}
B_1=2(\lambda^{-T}-\lambda)^{-1}=2(1-\lambda^T\lambda)^{-1}\lambda^T=2\lambda^T(1-\lambda\lambda^T)^{-1}    
\end{equation}
Using the constraints (\ref{skeleton condition}), the other matrices can be obtained by solving,
\begin{equation}\label{A_2}
\begin{aligned}
A_2&=-2(\lambda^{-1}-\lambda^T)^{-1}=-2\lambda(1-\lambda^T\lambda)^{-1}=-2(1-\lambda\lambda^T)^{-1}\lambda\\
B_2&B_2=1-B_1A_2~~~~~~~~~~~~A_1A_1=1-A_2B_1
\end{aligned}    
\end{equation}
The result is,
\begin{equation}\label{A and B}
\begin{aligned}
B_2&=-\bigg[(1-\lambda^T\lambda)^{-1}+(\lambda^{-1}\lambda^{-T}-1)^{-1}\bigg]\\
A_1&=\bigg[(1-\lambda\lambda^T)^{-1}+(\lambda^{-T}\lambda^{-1}-1)^{-1}\bigg]    
\end{aligned}
\end{equation}
This results in the $\mathcal{E}_\lambda-$map to be,
\begin{equation}
\begin{aligned}
\mathcal{E}_\lambda(\{X,Y\})=&\{\big[\left(1-\lambda \lambda^{T}\right)^{-1}+\left(\lambda^{-T} \lambda^{-1}-1\right)^{-1}\big]X-2\left(\lambda^{-1}-\lambda^{T}\right)^{-1}Y,\\
&2\left(\lambda^{-T}-\lambda\right)^{-1} X-\big[\left(1-\lambda^{T} \lambda\right)^{-1}+\left(\lambda^{-1} \lambda^{-T}-1\right)^{-1}\big]Y\} 
\end{aligned}
\end{equation}
This choice for matrix $B_1$ (\ref{B_1}) is motivated by the known form of $\mathcal{E}_\lambda-$map  considered in \cite{demulder2019doubled} for the case when $\lambda$ is assumed to be a number. The matrices $A_i$ and $B_i$ agree with the mapping used in \cite{demulder2019doubled} as they simplify as following for $\lambda$ being a number,
\begin{equation}
A_1=-B_2=\frac{1+\lambda^2}{1-\lambda^2}~~~~~~~~~~~~~~~-A_2=B_1=\frac{2\lambda}{1-\lambda^2}    
\end{equation}
The $\mathcal{E}_\lambda-$map is related to $\mathcal{E}:\mathcal{D}\rightarrow\mathcal{D}$ via the isomorphism $\Phi_\lambda:\mathcal{D}\rightarrow\mathcal{G}\oplus\mathcal{G}$ we obtain the constraints,
\begin{equation}\label{phi E relation}
\mathcal{E}_{\lambda} \circ \Phi_{\lambda}=\Phi_{\lambda} \circ \mathcal{E},~~~~~~~~~~~~~\text{where,}~~~~~~~~~~~~~\mathcal{E}(x_1\dot{+}{x_2})=(x_2\dot{+}x_1{}) 
\end{equation}
We now determine this isomorphism $\Phi_\lambda$ using the structure considered for $\mathcal{E}_\lambda-$map in (\ref{E skeleton}). Let us also assume a general form for $\Phi_\lambda$ for $\alpha\dot{+}\beta\in\mathcal{D}$, where we shall use the known conditions (\ref{skeleton condition}) to find the $C_i$ and $D_i$
\begin{equation}
\Phi_\epsilon(\alpha\dot{+}{\beta})=\{C_1\alpha+C_2\beta,D_1\alpha+D_2\beta\}    
\end{equation}
Substituting the above and the form of $\mathcal{E}_\lambda$-map in the constraint (\ref{phi E relation}), we obtain the following relations,
\begin{equation}\label{other cond}
\begin{aligned}
&A_1C_1+A_2D_1=C_2~~~~~~~~~~~~~~A_1C_2+A_2D_2=C_1\\
&B_1C_1+B_2D_1=D_2~~~~~~~~~~~~~~B_1C_2+B_2D_2=D_1
\end{aligned}
\end{equation}
Further demanding that the inner product with respect to $\mathcal{D}$ is preserved upto a constant factor ($\Lambda$) by the isomorphism $\Phi_\lambda$ we obtain the constraints,
\begin{equation}\label{inner product preserved}
\left(\Phi_{\lambda}(x), \Phi_{\lambda}(y)\right)_{\mathcal{G} \oplus \mathcal{G}}=\Lambda(x, y)_{\mathcal{D}}, \quad x, y \in \mathcal{D}
\end{equation}
\begin{equation}\label{CD other cond}
\begin{aligned}
&C^T_1C_1-D^T_1D_1=0~~~~~~~~~~C^T_2C_2-D^T_2D_2=0\\
&C_1^TC_2-D^T_1D_2=C^T_2C_1-D^T_2D_1=\Lambda
\end{aligned}    
\end{equation}
Choosing $C_1=1$ and $D_1=1$, we determine $C_2$ and $D_2$ using the conditions in (\ref{other cond}) as,
\begin{equation}
\begin{aligned}
C_2=A_1+A_2=&(1-\lambda\lambda^T)^{-1}+(\lambda^{-T}\lambda^{-1}-1)^{-1}-2(\lambda^{-1}-\lambda^T)^{-1}\\
=&(1-\lambda\lambda^T)^{-1}\big[1+\lambda\lambda^T-2\lambda\big]\\
D_2=B_1+B_2=&-(1-\lambda^T\lambda)^{-1}-(\lambda^{-1}\lambda^{-T}-1)^{-1}+2(\lambda^{-T}-\lambda)^{-1}\\
=&-(1-\lambda^T\lambda)^{-1}\big[1+\lambda^T\lambda-2\lambda^T\big]
\end{aligned}
\end{equation}
These choices of $C_i$ and $D_i$ satisfy all the above mentioned conditions results in the isomorphism $\Phi_\lambda$ for $(x-1\dot{+}x_2)\in\mathcal{D}$,
\begin{equation}
\Phi_\epsilon(x_1\dot{+}x_2)=\{x_1+(1-\lambda\lambda^T)^{-1}\big[1+\lambda\lambda^T-2\lambda\big]x_2,x_1-(1-\lambda^T\lambda)^{-1}\big[1+\lambda^T\lambda-2\lambda^T\big]x_2\}    
\end{equation}
which respecting consistency also boils down to the case considered in \cite{demulder2019doubled} for when $\lambda$ is simply a number,
\begin{equation}
\Phi_\epsilon(x_1\dot{+}x_2)=\bigg\{x_1+\frac{1-\lambda}{1+\lambda}x_2,x_1-\frac{1-\lambda}{1+\lambda}x_2\bigg\}~~~~~~~~\text{where,}~~\frac{1-\lambda}{1+\lambda}=\epsilon    
\end{equation}
As a check on the procedure, we can look at the conditions in (\ref{CD other cond}) for $C_i$ and $D_i$ when written in terms of $A_i$ and $B_i$ and the assumed values,
\begin{equation}
\begin{aligned}
&C^T_1C_1-D^T_1D_1=1-1=0\\
&C_2-D_2=C^T_2-D^T_2=(1-\lambda\lambda^T)^{-1}+(\lambda^{-T}\lambda^{-1}-1)^{-1}-2(\lambda^{-1}-\lambda^T)^{-1}\\
&~~~~~~~~~~~~~~~~~~~~~~~~~~~~+(1-\lambda^T\lambda)^{-1}+(\lambda^{-1}\lambda^{-T}-1)^{-1}-2(\lambda^{-T}-\lambda)^{-1}\\
&C_2^TC_2=D^T_2D_2=(A_1^T+A^T_2)(A_1+A_2)=(B_1^T+B^T_2)(B_1+B_2)
\end{aligned}    
\end{equation}
The final condition above is satisfied only when we assume the constraints obtained on $A_i$ and $B_i$ in (\ref{skeleton condition}) providing a good consistency check. To derive the generalized metric $\mathcal{H}_{AB}$, we look at the following relation where $\mathcal{E}_\lambda-$map is written as a matrix $\mathcal{E}_A{}^B$,
\begin{equation}
\mathcal{E}_A{}^B=\mathcal{H}_{AC}\eta^{CB}    
\end{equation}
The matrix $\mathcal{E}_{A}{}^B$ is obtained by using the inner product $(\mathbb{T}_A,\mathbb{T}_B)_{\mathcal{D}}=\eta_{AB}$, where the basis is formed by the diagonal and anti-diagonal generators $x\to\frac{\{x,x\}}{\sqrt{2}}$ and $x\to\frac{\{x,-x\}}{\sqrt{2}}$ respectively along with,
\begin{equation}
\eta_{CB}=\begin{pmatrix}
0&\delta_c{}^b\\
\delta^c{}_b&0
\end{pmatrix}    
\end{equation}
This results in the following generalized metric $\mathcal{H}_{AB}$ where $A_i$ and $B_i$ are as defined earlier in (\ref{B_1}), (\ref{A_2}) and (\ref{A and B})
\begin{equation}
\mathcal{H}_{A B}=\frac{1}{2}\left(\begin{array}{ll}
\left(A_{1}-A_{2}+B_{1}-B_{2}\right)\kappa^{-1} & \left(A_{1}+A_{2}+B_{1}+B_{2}\right)\kappa^{-1} \\
\left(A_{1}-A_{2}-B_{1}+B_{2}\right)\kappa & \left(A_{1}+A_{2}-B_{1}-B_{2}\right)\kappa
\end{array}\right)
\end{equation}
This generalized metric is acted on by the generalized frame fields $E_A{}^I$ (\ref{frame field}) to obtain the coordinate dependent generalized metric $\hat{\mathcal{H}}_{\hat{I}\hat{J}}$. The information of the embedded theory is further read off from this metric $\hat{\mathcal{H}}_{\hat{I}\hat{J}}$. 
\begin{equation}
\hat{\mathcal{H}}^{\hat{I}\hat{J}}=\big(E_A{}^{\hat{I}}\big)^T\mathcal{H}^{AB}E_B{}^{\hat{J}}
\end{equation}

\vspace{2.5cm}
\bibliographystyle{ieeetr}

\begin{thebibliography}{10}

\bibitem{Tseytlin1990DualitySF}
A.~A. Tseytlin, ``Duality symmetric formulation of string world sheet
  dynamics,'' {\em Physics Letters B}, vol.~242, pp.~163--174, 1990.

\bibitem{2009}
C.~Hull and B.~Zwiebach, ``Double field theory,'' {\em Journal of High Energy
  Physics}, vol.~2009, p.~099–099, Sep
  {2009}\href{https://arxiv.org/abs/0904.4664}{ ~arXiv:0904.4664 [hep-th]}.

\bibitem{hull1003}
O.~Hohm, C.~Hull, and B.~Zwiebach, ``Background independent action for double
  field theory,'' {\em Journal of High Energy Physics}, vol.~2010, Jul
  {2010}\href{https://arxiv.org/abs/1003.5027}{~~ arXiv:1003.5027 [hep-th]}.

\bibitem{1995}
C.~Klimčík and P.~Ševera, ``Dual non-abelian duality and the drinfeld
  double,'' {\em Physics Letters B}, vol.~351, p.~455–462, Jun
  {1995}\href{https://arxiv.org/abs/hep-th/9502122}{ ~arXiv:hep-th/9502122}.

\bibitem{1996}
C.~Klimčík and P.~Ševera, ``Poisson-lie ${T}$-duality and loop groups of
  drinfeld doubles,'' {\em Physics Letters B}, vol.~372, p.~65–71, Apr
  {1996}\href{https://arxiv.org/abs/hep-th/9512040}{~ arXiv:hep-th/9512040}.

\bibitem{Hull_2005}
C.~Hull, ``A Geometry for non-geometric string backgrounds,'' {\em Journal of High Energy Physics}, vol.~2005,
  no.~10, p.~065, {2005}\href{https://arxiv.org/abs/hep-th/0406102}{{
  arXiv:hep-th/0406102v3}}.

\bibitem{hull2009non}
C.~Hull and R.~Reid-Edwards, ``Non-geometric backgrounds, doubled geometry and
  generalised ${T}$-duality,'' {\em Journal of High Energy Physics}, vol.~2009,
  no.~09, p.~014, {2009}\href{https://arxiv.org/abs/0902.4032}{{
  arXiv:0902.4032v1 [hep-th]]}}.
  
\bibitem{Andriot_2012}
D.~Andriot, O.~Hohm, M.~Larfors, D.~Lüst, and P.~Patalong, ``Non-geometric fluxes in supergravity and double field theory,'' {\em Fortsch. Phys.}, vol.~60, p.~1150--1186, {2012}\href{https://arxiv.org/abs/1204.1979}{{
 arXiv:1204.1979 [hep-th]}}.  

\bibitem{demulder2019doubled}
S.~Demulder, F.~Hassler, and D.~C. Thompson, ``Doubled aspects of generalised
  dualities and integrable deformations,'' {\em Journal of High Energy
  Physics}, vol.~2019, no.~2, pp.~1--55,
  {2019}\href{https://arxiv.org/abs/1810.11446}{{ arXiv:1810.11446 [hep-th]}}.

\bibitem{1402.2105}
C.~Klim{\v{c}}{\'\i}k, ``Integrability of the bi-Yang-Baxter sigma-model,'' {\em Lett. Math. Phys.}, vol.~104, pp.~1095--1106,
  {2014}\href{https://arxiv.org/abs/1402.2105}{{ arXiv:1402.2105v1 [math-ph]}}.

\bibitem{1606.03016}
C.~Klim{\v{c}}{\'\i}k, ``Poisson\textendash{}Lie T-duals of the bi-Yang\textendash{}Baxter models",
,'' {\em Phys. Lett. B}, vol.~760, pp.~345--349,
  {2016}\href{https://arxiv.org/abs/1606.03016}{{ arXiv:1606.03016v2 [hep-th]}}.

\bibitem{17new}
F.~Hassler, ``Poisson-lie ${T}$-duality in double field theory,'' {\em Physics
  Letters B}, vol.~807, Aug {2020}\href{https://arxiv.org/abs/1707.08624}{~ arXiv:1707.08624 [hep-th]}.

\bibitem{klimvcik2015eta}
C.~Klim{\v{c}}{\'\i}k, ``$\eta$ and $\lambda$ deformations as
  $\mathcal{E}$-models,'' {\em Nuclear Physics B}, vol.~900, pp.~259--272,
  {2015}\href{https://arxiv.org/abs/1508.05832}{{ arXiv:1508.05832v2
  [hep-th]}}.

\bibitem{demulder2019invitation}
S.~Demulder, F.~Hassler, and D.~C. Thompson, ``An invitation to
  $\text{Poisson-Lie T-}$duality in $\text{Double Field Theory}$ and its
  applications,''
  {2019}\href{https://arxiv.org/abs/1904.09992}{~arXiv:1904.09992 [hep-th]}.

\bibitem{thompson2019introduction}
D.~C. Thompson, ``An introduction to generalised dualities and their
  applications to holography and integrability,''
  {2019}\href{https://arxiv.org/abs/1904.11561}{~ arXiv:1904.11561 [hep-th]}.

\bibitem{borsato2021supergravity}
R.~Borsato and S.~Driezen, ``Supergravity solution-generating techniques and
  canonical transformations of $\sigma$-models from ${O (D, D)}$,'' {\em
  Journal of High Energy Physics}, vol.~2021, no.~5, pp.~1--69,
  {2021}\href{https://arxiv.org/abs/2102.04498}{{ arXiv:2102.04498v2
  [hep-th]}}.

\bibitem{21new}
R.~Borsato, S.~Driezen, and F.~Hassler, ``An algebraic classification of
  solution generating techniques,'' {\em Physics Letters B}, vol.~823, Dec
  {2021}\href{https://arxiv.org/abs/2109.06185}{~~ arXiv:2109.06185 [hep-th]}.

\bibitem{2014}
K.~Sfetsos, ``Integrable interpolations: From exact $\text{CFT}$s to
  non-abelian ${T}$-duals,'' {\em Nuclear Physics B}, vol.~880, p.~225–246,
  Mar {2014}\href{https://arxiv.org/abs/1312.4560}{~ arXiv:1312.4560 [hep-th]}.

\bibitem{sfetsos2015generalised}
K.~Sfetsos, K.~Siampos, and D.~C. Thompson, ``{Generalised integrable
  $\lambda$-and $\eta$-deformations and their relation},'' {\em Nuclear Physics
  B}, vol.~899, pp.~489--512, {2015}\href{https://arxiv.org/abs/1506.05784}{{
  arXiv:1506.05784v3 [hep-th]}}.

\bibitem{20212}
G.~Itsios, K.~Sfetsos, and K.~Siampos, ``Novel integrable interpolations,''
  {\em Nuclear Physics B}, vol.~971, p.~115515, Oct 2021.

\bibitem{demulder2015integrable}
S.~Demulder, K.~Sfetsos, and D.~C. Thompson, ``Integrable
  $\lambda$-deformations: Squashing coset $\text{CFT}$s and ${AdS_5\times
  S^5}$,'' {2015}\href{https://arxiv.org/abs/1504.02781}{~arXiv:1504.02781
  [hep-th]}.

\bibitem{2019}
S.~Driezen, A.~Sevrin, and D.~C. Thompson, ``Integrable asymmetric
  $\lambda$-deformations,'' {\em Journal of High Energy Physics}, vol.~2019,
  Apr {2019}\href{https://arxiv.org/abs/1902.04142}{~ arXiv:1902.04142
  [hep-th]}.

\bibitem{2021}
S.~Driezen and K.~Sfetsos, ``Integrable $\lambda$-deformations of the
  $\text{Euclidean}$ black string,'' {\em Nuclear Physics B}, vol.~964,
  p.~115327, Mar
  {2021}\href{https://arxiv.org/abs/2012.08527}{~arXiv:2012.08527 [hep-th]}.

\bibitem{2016}
Y.~Chervonyi and O.~Lunin, ``Generalized $\lambda$-deformations of
  $\text{AdS}_p \times \text{S}^q$,'' {\em Nuclear Physics B}, vol.~913,
  p.~912–941, Dec
  {2016}\href{https://arxiv.org/abs/1608.06641}{~arXiv:1608.06641 [hep-th]}.

\bibitem{2004}
I.~Bena, J.~Polchinski, and R.~Roiban, ``Hidden symmetries of the
  $\text{AdS}_5\times \text{S}^5$ superstring,'' {\em Physical Review D},
  vol.~69, Feb {2004}\href{https://arxiv.org/abs/hep-th/0305116}{~
  arXiv:hep-th/0305116}.

\bibitem{20091}
C.~Klimčík, ``On integrability of the $\text{Yang–Baxter}$
  $\sigma$-model,'' {\em Journal of Mathematical Physics}, vol.~50, p.~043508,
  Apr {2009}\href{https://arxiv.org/abs/0802.3518}{~arXiv:0802.3518 [hep-th]}.

\bibitem{2018}
O.~Lunin and W.~Tian, ``Analytical structure of the generalized
  $\lambda$-deformation,'' {\em Nuclear Physics B}, vol.~929, p.~330–352, Apr
  {2018}\href{https://arxiv.org/abs/1711.02735}{~ arXiv:1711.02735 [hep-th]}.

\bibitem{20161}
Y.~Chervonyi and O.~Lunin, ``Supergravity background of the $\lambda$-deformed
  $\text{AdS}_3\times \text{S}^3$ supercoset,'' {\em Nuclear Physics B},
  vol.~910, p.~685–711, Sep
  {2016}\href{https://arxiv.org/abs/1606.00394}{~arXiv:1606.00394 [hep-th]}.

\bibitem{20141}
K.~Sfetsos and D.~C. Thompson, ``Spacetimes for $\lambda$-deformations,'' {\em
  Journal of High Energy Physics}, vol.~2014, Dec
  {2014}\href{https://arxiv.org/abs/1410.1886}{~ arXiv:1410.1886 [hep-th]}.

\bibitem{20131}
G.~Aldazabal, D.~Marqués, and C.~Núñez, ``Double field theory: a pedagogical
  review,'' {\em Classical and Quantum Gravity}, vol.~30, p.~163001, Jul
  {2013}\href{https://arxiv.org/abs/1305.1907}{~ arXiv:1305.1907 [hep-th]}.

\bibitem{20192}
Y.~Sakatani, ``Type $\text{II DFT}$ solutions from $\text{Poisson–Lie}$
  ${T}$-duality/plurality,'' {\em Progress of Theoretical and Experimental
  Physics}, vol.~2019, Jul {2019}\href{https://arxiv.org/abs/1903.12175}{~
  arXiv:1903.12175 [hep-th]}.


\end{thebibliography}

\end{document}